%% file: sltnf.tex
\documentclass[12pt]{article}
\usepackage{amsfonts}
\usepackage{color}

\newtheorem{definition}{Definition}[section]
\newtheorem{example}{Example}[section]

\newtheorem{lemma}{Lemma}[section]
\newtheorem{theorem}[lemma]{Theorem}
\newtheorem{corollary}[lemma]{Corollary}

\begin{document}

\title{\Large \bf Enhancing Global SLS-Resolution with Loop Cutting and Tabling Mechanisms}

\author{Yi-Dong Shen\\
{\small  Laboratory of Computer Science, Institute of Software}\\
{\small Chinese Academy of Sciences, Beijing 100080, China}\\
{\small Email: ydshen@ios.ac.cn}\\[.1in]
Jia-Huai You and Li-Yan Yuan\\
{\small  Department of Computing Science, University of Alberta}\\
{\small  Edmonton, Alberta, Canada T6G 2H1}\\
{\small  Email: \{you, yuan\}@cs.ualberta.ca}
}

\date{}

\maketitle
 
\begin{abstract} 
Global SLS-resolution is a well-known procedural semantics for top-down computation
of queries under the well-founded model. It inherits from SLDNF-resolution 
the {\em linearity} property of derivations, which makes it easy and efficient to implement
using a simple stack-based memory structure. However, like
SLDNF-resolution it suffers from the problem of infinite loops and redundant computations. 
To resolve this problem, in this paper we develop a new procedural semantics, 
called {\em SLTNF-resolution}, by enhancing 
Global SLS-resolution with loop cutting and tabling mechanisms. SLTNF-resolution
is sound and complete w.r.t. the well-founded semantics 
for logic programs with the bounded-term-size property, and
is superior to existing linear tabling procedural semantics such as SLT-resolution.\\[.1in]
\end{abstract} 
 
\section{Introduction}
There are two types of semantics for a logic program: a declarative semantics and
a procedural semantics. The declarative semantics formally defines the meaning 
of a logic program by specifying an {\em intended model} among all models of the
logic program, whereas the procedural semantics implements/computes the declarative semantics 
by providing an algorithm for evaluating queries against the logic program. Most 
existing procedural semantics are built upon the well-known {\em resolution rule}
created by Robinson \cite{Robinson65}. 

Prolog is the first yet the most popular logic programming language \cite{Kow74}.
It adopts {\em SLDNF-resolution} as its procedural semantics \cite{clark78}. 
One of the best-known properties of SLDNF-resolution is its {\em linearity} of derivations,
i.e., its query evaluation (i.e., SLDNF-derivations) 
constitutes a search tree, called an {\em SLDNF-tree}, which can be implemented
easily and efficiently using a simple stack-based memory structure \cite{WAM83, ZHOU96}. 
However, SLDNF-resolution suffers from two serious problems. First, 
its corresponding declarative semantics,
i.e. the {\em predicate completion semantics} \cite{clark78}, is based on two truth values 
(either {\em true} or {\em false}) and thus incurs inconsistency for some logic programs like 
$P=\{p(a) \leftarrow \neg p(a)\}$ \cite{Ld87, sheph88}. 
Second, it may generate infinite loops and a large amount
of redundant sub-derivations \cite{BAK91, DD93, shen-tocl}.

To overcome the first problem with SLDNF-resolution, 
the {\em well-founded semantics} \cite{VRS91} 
is introduced as an alternative to the predicate completion semantics.
A well-founded model accommodates three truth values: {\em true, false} and {\em undefined}, so that
inconsistency is avoided by letting atoms that are recursively
connected through negation undefined. Several procedural semantics have been developed
as an alternative to SLDNF-resolution
to compute the well-founded semantics, among the most representative of which are
{\em Global SLS-resolution} \cite{Prz89,Ross92} 
and {\em SLG-resolution} \cite{CSW95,chen96,BD98}.

Global SLS-resolution is a direct extension of SLDNF-resolution. It evaluates queries
under the well-founded semantics by
generating a search tree, called an {\em SLS-tree}, in the same way as 
SLDNF-resolution does except that infinite derivations are treated 
as {\em failed} and infinite recursions through negation as {\em undefined}.
Global SLS-resolution retains the linearity property of SLDNF-resolution,  
but it also inherits the problem of infinite loops
and redundant computations. Moreover, Global SLS-resolution handles 
negation as follows: A ground atom $A$ is false
when all branches of the SLS-tree for $A$ are either 
infinite or end at a failure leaf. Infinite branches make 
Global SLS-resolution not effective in general \cite{Ross92}.  

To resolve infinite loops and redundant computations, the {\em tabling} technique
is introduced \cite{TS86,war92}. The main idea of tabling is to store intermediate answers
of subgoals and then apply them to solve variants of the subgoals. 
With tabling no variant subgoals will be recomputed by applying
the same set of clauses, so infinite loops can be avoided and 
redundant computations be substantially reduced.
There are two typical ways to make use of tabling to compute the well-founded semantics.
One is to directly enhance SLDNF-resolution
or Global SLS-resolution with tabling while the other
is to create a new tabling mechanism with a different derivation structure. 
SLG-resolution results from the second way \cite{BD98,chen96}. Due to the use of tabling,
SLG-resolution gets rid of infinite loops and reduces redundant computations. However,
it does not have the linearity property since its query
evaluation constitutes a search forest instead of a search tree. 
As a result, it cannot be implemented in the same way as SLDNF-resolution using a simple stack-based 
memory structure \cite{SSW94,xsb98,swift99}. 

In \cite{shen2002} an attempt is made to directly enhance SLDNF-resolution with tabling
to compute the well-founded semantics, which leads to 
a tabling mechanism, called {\em SLT-resolution}. SLT-resolution
retains the linearity property, thus is referred to as a 
{\em linear tabling} mechanism. Due to the use of tabling, it is free of infinite
loops and has fewer redundant computations than SLDNF-resolution. However, 
SLT-resolution has the following two major drawbacks:
(1) It defines positive loops and negative loops 
based on the same ancestor-descendant relation, which makes
loop detection and handling quite costly since a loop may go across several 
(subsidiary) SLT-trees. (2) It makes use of answer iteration to 
derive all answers of looping subgoals, but provides no answer completion criteria
for pruning redundant derivations. Note that answer completion is the key
to an efficient tabling mechanism.

In this paper, we develop a new procedural semantics, called {\em SLTNF-resolution},
for the well-founded semantics by enhancing Global SLS-resolution 
with tabling and loop cutting mechanisms. SLTNF-resolution retains the linearity
property and makes use of tabling to get rid of all loops and reduce redundant computations.
It defines positive and negative loops in terms of two different ancestor-descendant
relations, one on subgoals within an SLS-tree and the other on SLS-trees, so that
positive and negative loops can be efficiently detected and handled.
It employs two effective criteria for answer completion of tabled subgoals
so that redundant derivations can be pruned as early as possible.
All these mechanisms are integrated into an algorithm quite like
that for generating SLS-trees.

The paper is organized as follows. Section 2 reviews Global SLS-resolution.
Section 3 defines ancestor-descendant relations for identifying positive and
negative loops, develops an algorithm for generating SLTNF-trees, establishes
criteria for determining answer completion of tabled subgoals, and proves 
the correctness of SLTNF-resolution. Section 4 mentions some related work, 
and Section 5 concludes.

\section{Preliminaries and Global SLS-Resolution} 
\label{sec2}
In this section, we review some standard terminology of
logic programs \cite{Ld87} and recall the definition of Global SLS-Resolution. 
We do not repeat the definition of the well-founded model here; 
it can be found in \cite{VRS91,Prz89,Prz90}
and many other papers.

Variables begin with a capital letter, and predicate, function and constant symbols with a 
lower case letter. By a {\em variant} of a literal $L$ we mean a literal $L'$ that is identical to 
$L$ up to variable renaming.

\begin{definition}
{\em
A {\em general logic program} (logic program for short) is a finite set
of clauses of the form

$\qquad A\leftarrow L_1,..., L_n$

\noindent where $A$ is an atom and $L_i$s are literals. 
$A$ is called the {\em head} and $L_1,...,L_n$ is called the
{\em body} of the clause. When $n=0$, the ``$\leftarrow$'' symbol is omitted. 
If a logic program has no clause with negative
literals in its body, it is called a {\em positive logic program}.
}
\end{definition}

\begin{definition}
{\em
A {\em goal} $G$ is a headless clause
$\leftarrow L_1,..., L_n$ where each $L_i$ is called a {\em subgoal}.
A goal is also written as $G = \leftarrow Q$ where $Q = L_1,..., L_n$ is called a {\em query}.
A {\em computation rule} (or {\em selection rule}) is a rule for selecting one subgoal from a goal.
}
\end{definition}

Let $G_i=\leftarrow L_1,..., L_j,..., L_n$ be a 
goal with $L_j$ a positive subgoal. Let $C = L\leftarrow F_1,...,F_m$ be 
a clause such that $L$ and $L_j$ are unifiable, i.e. $L\theta = L_j\theta$ where $\theta$ is an mgu (most general unifier). 
The {\em resolvent} of $G_i$ and $C$ on $L_j$ is a goal
$G_k=\leftarrow (L_1,...,L_{j-1}, F_1,...,F_m, L_{j+1},..., L_n)\theta$.
In this case, we say that the proof of $G_i$ is reduced to the proof of $G_k$.

The initial goal, $G_0=\leftarrow L_1,..., L_n$, is called
a {\em top} goal. Without loss of generality, we shall assume throughout
the paper that a top goal consists only of one atom (i.e. $n=1$ and $L_1$
is a positive literal). 

Trees are used to depict the search space of a top-down query evaluation procedure. 
For convenience, a node in such a tree is represented by $N_i:G_i$ where $N_i$ is the 
node name and $G_i$ is a goal labeling the node. Assume no two 
nodes have the same name, so we can refer to nodes by their names.

Let $P$ be a logic program and $G_0 = \leftarrow Q$ a top goal. Global SLS-resolution is the 
process of constructing SLS-derivations from $P\cup \{G_0\}$ via a computation rule $R$. 
An {\em SLS-derivation} is a partial branch beginning at the root $N_0:G_0$ of an SLS-tree. 
Every leaf of an SLS-tree is either a {\em success} leaf or a {\em failure} leaf or a {\em flounder}
leaf or an {\em undefined} leaf.\footnote{In \cite{Prz92}, 
an undefined leaf is called a {\em non-labeled} leaf.}
$Q$ is a {\em non-floundering query} if no SLS-tree for evaluating $Q$ under 
$R$ contains a flounder leaf.

An SLS-tree is {\em successful} if it has a success leaf. It is {\em failed}
if all of its branches are either infinite or end at a failure leaf. It is {\em floundered} if it 
contains a floundered leaf and is not successful. An SLS-tree is {\em undefined} if it is neither
successful nor failed nor floundered.

There are two slightly different definitions of an SLS-tree: Przymusinski's definition \cite{Prz89,Prz92}
and Ross' definition \cite{Ross92}. Przymusinski's definition requires a level mapping (called {\em strata})
to be associated with literals and goals, while Ross' definition requires the computation rule
to be {\em preferential}, i.e. positive subgoals are selected ahead of negative ones and negative
subgoals are selected in parallel. Both of the two definitions allow infinite branches and infinite
recursion through negation. The following definition of an SLS-tree is obtained by combining
the two definitions.

\begin{definition}[SLS-trees \cite{Prz89,Prz92,Ross92}]
\label{sls-tree}
{\em
Let $P$ be a logic program, $G_0$ a top goal, and $R$ a computation rule.
The {\em SLS-tree} $T_{N_0:G_0}$ for $P \cup \{G_0\}$ via $R$ is a tree 
rooted at $N_0:G_0$ such that for any node $N_i:G_i$ in the tree
with $G_i=\leftarrow L_1,...,L_n$:

\begin{enumerate}
\item
If $n=0$ then $N_i$ is a {\em success} leaf, marked by $\Box_t$.

\item
\label{l1}
If $L_j$ is a positive literal selected by $R$, then for each clause $C$ in $P$
whose head is unifiable with $L_j$, $N_i$ has a child $N_k:G_k$ where $G_k$
is the resolvent of $C$ and $G_i$ on $L_j$. If no such a clause exists in $P$, then
$N_i$ is a {\em failure} leaf, marked by $\Box_f$.

\item
\label{l2}
Let $L_j=\neg A$ be a negative literal selected by $R$. If $A$ is not ground then
$N_i$ is a {\em flounder} leaf, marked by $\Box_{fl}$, else let $T_{N_{i+1}:\leftarrow A}$ 
be an (subsidiary) SLS-tree for $P \cup \{\leftarrow A\}$ via $R$. We consider four cases:
\begin{enumerate}
\item
\label{la}
If $T_{N_{i+1}:\leftarrow A}$ is failed then $N_i$ has only one child that is labeled
by the goal $\leftarrow L_1,...,L_{j-1},L_{j+1},...,L_n$.

\item
\label{lb}
If $T_{N_{i+1}:\leftarrow A}$ is successful then $N_i$ is a {\em failure} leaf, marked by $\Box_f$.

\item
\label{lc}
If $T_{N_{i+1}:\leftarrow A}$ is floundered then $N_i$ is a {\em flounder} leaf, marked by $\Box_{fl}$.

\item
\label{ld}
Otherwise (i.e. $T_{N_{i+1}:\leftarrow A}$ is undefined), we mark $L_j$ in $G_i$ as {\em skipped}
and use the computation rule $R$ to select a new literal $L_k$ from $G_i$ and apply 
the resolution steps \ref{l1} and \ref{l2} to the goal $G_i$. If all literals in $G_i$ were
already marked as skipped then $N_i$ is an {\em undefined} leaf, marked by $\Box_u$.
\end{enumerate}
\end{enumerate} 
}
\end{definition}

We make two remarks. First, the level mapping/strata used in Przymusinski's definition
is implicit in Definition \ref{sls-tree}. That is, in case \ref{l2}
the level/stratum of $A$ is less than the level/stratum of $G_i$ if and only if
either case \ref{la} or case \ref{lb} or case \ref{lc} holds. Second, the preferential
restriction of Ross' definition to the computation rule is relaxed by marking undefined subgoals as skipped
and then continuing to select new subgoals from the remaining subgoals in $G_i$ for 
evaluation (see case \ref{ld}). A leaf is undefined if and only if all its subgoals are 
marked as skipped.

\begin{definition}
{\em
A {\em successful} (resp. {\em failed} or {\em undefined}) derivation for a goal $G$
is a branch beginning at the root of the SLS-tree for $G$ and ending at a success 
(resp. failure or undefined) leaf. A {\em correct answer substitution} for $G$ is the
substitution $\theta=\theta_1...\theta_n$, where $\theta_i$s are the most
general unifiers used at each step along the derivation, restricted to the variables
in $G$.
}
\end{definition}

It has been shown that Global SLS-resolution is sound and complete
with respect to the well-founded semantics for non-floundering
queries.

\begin{theorem}[\cite{Prz89,Prz92,Ross92}]
\label{sound-comp}
Let $P$ be a logic program, $R$ a computation rule,
and $G_0\leftarrow Q$ be a top goal with $Q$ a non-floundering
query under $R$. Let $WF(P)$ be the well-founded model of $P$.
\begin{enumerate}
\item
$WF(P) \models \exists (Q)$ if and only if the SLS-tree for $P \cup \{G_0\}$ via $R$ 
is successful.

\item
$WF(P) \models \forall (Q\theta)$ if and only if there exists a correct
answer substitution for $G_0$ more general than the substitution $\theta$.

\item
$WF(P) \models \neg \exists (Q)$ if and only if the SLS-tree for $P \cup \{G_0\}$ is failed.
\end{enumerate}
\end{theorem}

\begin{definition}
{\em
Let $N_i:G_i$ be a node in an SLS-tree $T_{N_r:G_r}$ where $A$ is the selected positive subgoal 
in $G_i$. The partial branches of $T_{N_r:G_r}$
beginning at $N_i$ that are used to evaluate $A$ constitute {\em sub-derivations}
for $A$. All such sub-derivations form a {\em sub-SLS-tree} for $A$ at $N_i$.
}
\end{definition}

By Theorem \ref{sound-comp}, for any correct answer substitution $\theta$ 
built from a successful sub-derivation for $A$,  
$WF(P) \models \forall (A\theta)$. 

Since Global SLS-resolution allows infinite derivations as well as
infinite recursion through negation, we may need 
infinite time to generate an SLS-tree. This is not feasible in practice.
In the next section, we resolve this problem by enhancing Global SLS-resolution
with both loop cutting and tabling mechanisms.

\section{SLTNF-Resolution}

We first define an ancestor-descendant relation on selected subgoals
in an SLS-tree. Informally, $A$ is an {\em ancestor subgoal} of $B$ if the proof of $A$ 
depends on (or in other words goes via) the proof of $B$. For example, 
let $M:\leftarrow A,A_1,...,A_m$ be a node in an SLS-tree, and 
$N:\leftarrow B_1\theta,...,B_n\theta, A_1\theta,...,A_m\theta$ be a child node of $M$ that 
is generated by resolving $M$ on the subgoal $A$ with a clause  
$A'\leftarrow B_1,...,B_n$ where $A\theta = A'\theta$. Then $A$ at $M$ is an ancestor subgoal 
of all $B_i\theta$s at $N$. However, such relationship does not 
exist between $A$ at $M$ and any $A_j\theta$ at $N$. It is easily seen 
that all $B_i\theta$s at $N$ inherit the ancestor subgoals of $A$ at $M$,
and that each $A_j\theta$ at $N$ inherits the ancestor subgoals of $A_j$ at $M$.
Note that subgoals at the root of an SLS-tree have no ancestor subgoals. 
 
Let $N_i:G_i$ and $N_k:G_k$ be two nodes and 
$A$ and $B$ be the selected subgoals in $G_i$ and $G_k$, respectively. 
When $A$ is an ancestor subgoal of $B$, we refer to $B$ as a {\em descendant 
subgoal} of $A$, $N_i$ as an {\em ancestor node} of $N_k$, and $N_k$ as  
a {\em descendant node} of $N_i$.
Particularly, if $A$ is both an ancestor subgoal and a variant,
i.e. an {\em ancestor variant subgoal}, of $B$,
we say the derivation goes into a {\em loop}, where $N_i$ and $N_k$ 
are respectively  called an {\em ancestor loop node} and a {\em descendant loop node},
and $A$ (at $N_i$) and $B$ (at $N_k$) are respectively  called an {\em ancestor loop 
subgoal} and a {\em descendant loop subgoal}.

The above ancestor-descendant relation is defined over subgoals and will be applied to detect
positive loops, i.e. loops within an SLS-tree. In order to handle negative loops 
(i.e. loops through negation like $A\leftarrow \neg B$ and $B\leftarrow \neg A$)
which occur across SLS-trees, we define an ancestor-descendant relation on SLS-trees. 
Let $N_i:\leftarrow \neg A, ...$ be a node in $T_{N_r:G_r}$,
with $\neg A$ the selected subgoal, and let $T_{N_{i+1}:\leftarrow A}$ 
be an (subsidiary) SLS-tree for $P \cup \{\leftarrow A\}$ via $R$.
$T_{N_r:G_r}$ is called an {\em ancestor SLS-tree} of $T_{N_{i+1}:\leftarrow A}$,
while $T_{N_{i+1}:\leftarrow A}$ is called a {\em descendant SLS-tree} of $T_{N_r:G_r}$.
Of course, the ancestor-descendant relation is transitive.

A negative loop occurs if an SLS-tree has a descendant SLS-tree, with the same goal at their roots.
For convenience, we use dotted edges to connect parent and child
SLS-trees, so that negative loops can be clearly identified.
Let $G_0$ be a top goal. We call $T_{N_0:G_0}$ together 
with all of its descendant SLS-trees
a {\em generalized SLS-tree}, denoted $GT_{P, G_0}$
(or simply $GT_{G_0}$ when no confusion would arise).
Therefore, a branch of a generalized SLS-tree may come 
across several SLS-trees through dotted edges. A 
{\em generalized SLS-derivation} is a partial branch beginning
at the root of a generalized SLS-tree.
  
Assume that all loops are detected and cut based on the ancestor-descendant relations.
This helps Global SLS-resolution get rid of infinite derivations and infinite
recursion through negation. However, applying such loop cutting mechanism alone
is not effective since some answers would be lost. In order to guarantee the
completeness of Global SLS-resolution with the loop cutting mechanism,
we introduce a tabling mechanism into SLS-derivations, leading to a tabulated SLS-resolution.

In tabulated resolutions, the set of predicate symbols in a logic program
is partitioned into two groups: {\em tabled predicate symbols} and
{\em non-tabled predicate symbols}. Subgoals with tabled predicate
symbols are then called {\em tabled subgoals}. A {\em dependency graph} \cite{ABW88}
is used to make such classification. Informally, for any predicate symbols $p$ and $q$,
there is an edge $p\rightarrow q$ in the dependency graph $G_P$ of a logic program $P$
if and only if $P$ contains a clause whose head contains $p$ and whose body contains $q$.
$p$ is a tabled predicate symbol if $G_P$ contains a cycle involving $p$. It is trivial to show 
that subgoals involved in any loops in SLS-trees must be tabled subgoals. 

Intermediate answers of tabled subgoals will be stored 
in tables once they are produced at some derivation stages. 
Such answers are called {\em tabled answers}.
For convenience of presentation, we organize a table into a compound
structure like $struct$ in pseudo $C^{++}$ language. That is, 
the table of an atom $A$, denoted $TB_A$, is internally an instance of 
the data type TABLE defined as follows:
\begin{tabbing}
$\quad$ \= typ\= edef struct \{\\
\>\> {\bf string} $\qquad$\= $atom$; //for $TB_A$, $atom = A$.\\
\>\> {\bf int} \> $comp$; //status of $atom$ indicating if all answers have been tabled.\\
\>\> {\bf set} \> $ans$; //tabled answers of $atom$.\\
\> \} TABLE;
\end{tabbing}

Answers of a tabled subgoal $A$ are stored in $TB_A\rightarrow ans$.
We say $TB_A$ is {\em complete} if $TB_A\rightarrow ans$
contains all answers of $A$. We use $TB_A\rightarrow comp = 1$ 
to mark the completeness of tabled answers. Clearly, the case
$TB_A\rightarrow comp = 1$ and $TB_A\rightarrow ans = \emptyset$
indicates that $A$ is false. 

We introduce a special subgoal, $u^*$, which is assumed to occur neither in
logic programs nor in top goals. $u^*$ will be used to substitute for
some ground negative subgoals whose truth values are 
temporarily undefined (i.e., whether they are true or false cannot be
determined at the current stage of derivation). We assume such a special subgoal will not
be selected by a computation rule. 

We also use a special subgoal, $LOOP$, to mark occurrence of a loop.

Augmenting SLS-trees with the loop cutting and tabling mechanisms leads to the  
following definition of SLTNF-trees. 
Here ``SLTNF'' stands for ``{\em L}inear {\em T}abulated resolution using
a {\em S}election/computation rule with {\em N}egation as Finite {\em F}ailure.'' 

\begin{definition}[SLTNF-trees]
\label{SLTNF-tree}
{\em
Let $P$ be a logic program, $G_0$ a top goal, and $R$ a computation rule.
Let $\cal T_P$ be a set of tables each of which contains a finite set of tabled answers.
The {\em SLTNF-tree} $T_{N_0:G_0}$ for $(P \cup \{G_0\},\cal T_P)$ via $R$ is a tree 
rooted at $N_0:G_0$ such that for any node $N_i:G_i$ in the tree
with $G_i=\leftarrow L_1,...,L_n$:

\begin{enumerate}
\item
\label{l20}
If $n=0$ then $N_i$ is a {\em success} leaf, marked by $\Box_t$, else
if $L_1=u^*$ then $N_i$ is a {\em temporarily undefined} leaf, marked by $\Box_{u^*}$, else
if $L_1=LOOP$ then $N_i$ is a {\em loop} leaf, marked by $\Box_{loop}$.

\item
If $L_j = p(.)$ is a positive literal selected by $R$, we consider two cases:

\begin{enumerate}

\item
If $TB_{L_j}\in \cal T_P$ with $TB_{L_j}\rightarrow comp=1$, then
for each tabled answer $A$ in $TB_{L_j}\rightarrow ans$, $N_i$ has a child node
$N_k:G_k$ where $G_k$ is the resolvent of $A$ and $G_i$ on $L_j$.
In case that $TB_{L_j}\rightarrow ans = \emptyset$, 
$N_i$ is a {\em failure} leaf, marked by $\Box_f$.

\item
\label{l4d}
Otherwise, for each tabled answer 
$A$ in $TB_{L_j}\rightarrow ans$ $N_i$ has a child node
$N_k:G_k$ where $G_k$ is the resolvent of $A$ and $G_i$ on $L_j$, and
\begin{enumerate}
\item
If $N_i$ is a descendant loop node then it has a child node $N_l:\leftarrow LOOP$.

\item
Otherwise, for each clause $C$ in $P$ whose head is unifiable
with $L_j$ $N_i$ has a child node $N_l:G_l$ where $G_l$ is the resolvent of $C$
and $G_i$ on $L_j$. If there are neither tabled answers nor clauses applicable to $N_i$ then 
$N_i$ is a {\em failure} leaf, marked by $\Box_f$.
\end{enumerate}
\end{enumerate}

\item
Let $L_j=\neg A$ be a negative literal selected by $R$. If $A$ is not ground then
$N_i$ is a {\em flounder} leaf, marked by $\Box_{fl}$, else we consider
the following cases:
\begin{enumerate}
\item
If $TB_{A}\in \cal T_P$ with $TB_{A}\rightarrow comp=1$ and
$TB_{A}\rightarrow ans = \emptyset$, then $N_i$ has only one child node $N_k:G_k$
with $G_k=\leftarrow L_1,...,L_{j-1},L_{j+1},...,L_n$.

\item
If $TB_{A}\in \cal T_P$ with $TB_{A}\rightarrow comp=1$ and
$TB_{A}\rightarrow ans = \{A\}$, then $N_i$ is a {\em failure} leaf, marked by $\Box_f$.

\item
\label{l5c}
Otherwise, if the current SLTNF-tree or one of its ancestor SLTNF-trees 
is with a goal $\leftarrow A$ at the root,
$N_i$ has only one child node $N_k:G_k$ where if $L_n\neq u^*$ then
$G_k = \leftarrow L_1,...,L_{j-1},L_{j+1},...,$ $L_n,u^*$ 
else $G_k = \leftarrow L_1,...,L_{j-1},L_{j+1},...,L_n$. 

\item
\label{l5d}
Otherwise, let $T_{N_r:\leftarrow A}$  be an (subsidiary) SLTNF-tree
for $(P \cup \{\leftarrow A\},\cal T_P)$ via $R$. 
We have the following cases:

\begin{enumerate}
\item
\label{l5d1}
If $T_{N_r:\leftarrow A}$ has a success leaf then 
$N_i$ is a {\em failure} leaf, marked by $\Box_f$.

\item
If $T_{N_r:\leftarrow A}$ has no success leaf but a flounder leaf then
$N_i$ is a {\em flounder} leaf, marked by $\Box_{fl}$.

\item
\label{l5d3-1}
({\em Negation As Finite Failure}) 
If all branches of $T_{N_r:\leftarrow A}$ end at either a failure or a loop
leaf where for each loop leaf generated from a descendant loop subgoal $V$,
no successful sub-derivation for its ancestor loop subgoal 
has a correct answer
substitution $\theta$ such that $V\theta$ is not in 
$\cal T_P$, then $N_i$ has only one child node $N_k:G_k$
with $G_k=\leftarrow L_1,...,L_{j-1},L_{j+1},...,L_n$.

\item
\label{l5d4}
Otherwise, $N_i$ has only one child node $N_k:G_k$ where if $L_n\neq u^*$
$G_k = \leftarrow L_1,...,L_{j-1},L_{j+1},...,$ $L_n,u^*$
else $G_k = \leftarrow L_1,...,L_{j-1},L_{j+1},...,L_n$. 
\end{enumerate}
\end{enumerate}
\end{enumerate} 
}
\end{definition}

Note that some commonly used concepts, such as derivations
(for goals), sub-derivations (for subgoals), sub-trees (for subgoals), generalized trees,
and correct answer substitutions, have the same meanings 
as in SLS-trees (see Section \ref{sec2}).

Positive loops are broken simply by disallowing descendant loop nodes
to apply clauses in $P$ for expansion (see case \ref{l4d}),
while negative loops are broken by substituting $u^*$ for
looping negative subgoals (see case \ref{l5c}). 
This guarantees that SLTNF-trees are finite for logic programs with
the bounded-term-size property (see Definition \ref{bounded-term-size}
and Theorem \ref{tree-finite}).

Note that $u^*$ is only introduced to signify existence
of subgoals whose truth values are temporarily non-determined
because of occurrence of positive or negative loops. So keeping
only one $u^*$ in a goal is enough for such a purpose. From 
case \ref{l20} of Definition \ref{SLTNF-tree} we see that goals
with $u^*$ cannot lead to a success leaf. However,
$u^*$ may well appear in a failure leaf since one of the other subgoals may fail
regardless of what truth values the temporarily undefined subgoals would take. 
This achieves the effect of what a preferential computation rule \cite{Ross92}
is supposed to achieve, although our computation rule is not necessarily preferential.

Observe that SLTNF-trees implement an {\em Negation As Finite Failure} (NAF) rule (see case \ref{l5d3-1}): 
A ground subgoal $\neg A$ fails if
$A$ succeeds, and succeeds if $A$ finitely fails after exhausting all answers of 
the loop subgoals involved in evaluating $A$. This NAF rule is the same as
that used in SLDNF-resolution \cite{clark78} except that loop leaves are considered.

The following example illustrates the process of constructing SLTNF-trees.

\begin{figure}[htb]  
\begin{center}
\input{fig2.latex}
\end{center}
\caption{The generalized SLTNF-tree $GT_{\leftarrow p(a,Y)}$ for 
$(P_1\cup \{\leftarrow p(a,Y)\},\emptyset)$.}\label{fig2}
\end{figure}
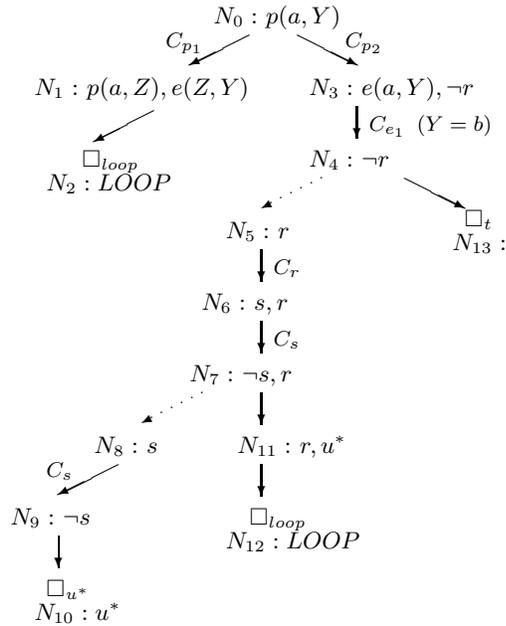

\begin{example}
\label{eg3-2}
{\em
Consider the following program and let $G_0=\leftarrow p(a,Y)$ be the top goal. 
\begin{tabbing} 
\hspace{.2in} $P_1$: \= $p(X,Y) \leftarrow p(X,Z),e(Z,Y).$ \`$C_{p_1}$ \\ 
\> $p(X,Y) \leftarrow e(X,Y), \neg r.$  \`$C_{p_2}$ \\ 
\> $e(a,b).$ \`$C_{e_1}$ \\ 
\> $e(b,c)$ \`$C_{e_2}$ \\ 
\> $r \leftarrow s, r.$ \`$C_r$ \\ 
\> $s \leftarrow \neg s.$ \`$C_s$
\end{tabbing} 
Let ${\cal T_P}_1=\emptyset$, and for convenience, let us choose the widely-used left-most computation rule
(i.e. we always select the left-most subgoal from a goal).
The generalized SLTNF-tree $GT_{\leftarrow p(a,Y)}$
for $(P_1\cup \{\leftarrow p(a,Y)\},\emptyset)$ is shown in Figure \ref{fig2},\footnote{
For simplicity, in depicting SLTNF-trees we omit the ``$\leftarrow$'' symbol in goals.} 
which consists of three finite SLTNF-trees that are rooted at $N_0$, $N_5$ and $N_8$, respectively.
Note that two positive loops are cut at $N_1$ and $N_{11}$, respectively,  
and one negative loop is cut at $N_9$.

$T_{N_5:\leftarrow r}$ has only one branch, which ends at a loop leaf $N_{12}$.
There is no successful sub-derivation for the ancestor loop subgoal $r$
at $N_5$, so the NAF rule is applicable. Thus, $\neg r$ at $N_4$ succeeds,
leading to a successful sub-derivation for $p(a,Y)$ at $N_0$ with a correct answer
substitution $\{Y/b\}$.
}
\end{example}

\begin{definition}[\cite{VG89}]
\label{bounded-term-size}
{\em
A logic program has the {\em bounded-term-size} property if there is a function
$f(n)$ such that whenever a top goal $G_0$ has no argument whose term size exceeds
$n$, then no subgoals and tabled answers in any generalized SLTNF-tree
$GT_{G_0}$ have an argument whose term size exceeds $f(n)$.
}
\end{definition}

The following result shows that the construction of SLTNF-trees
is always terminating for logic programs with the bounded-term-size property.

\begin{theorem}
\label{tree-finite}
Let $P$ be a logic program with the bounded-term-size property, 
$G_0$ a top goal and $R$ a computation rule. The generalized SLTNF-tree
$GT_{G_0}$ for $(P \cup \{G_0\},\cal T_P)$ via $R$ is finite.
\end{theorem}

\noindent {\bf Proof:}
First note that $GT_{G_0}$ contains no negative loops (see case \ref{l5c}).
The bounded-term-size property guarantees that no term occurring
on any path of $GT_{G_0}$ can have size greater than
$f(n)$, where $n$ is a bound on the size of terms in the top goal $G_0$.
Assume, on the contrary, that $GT_{G_0}$
is infinite. Since the branching factor of $GT_{G_0}$
(i.e. the average number of children of all nodes in the tree)
is bounded by the finite number of clauses in $P$,
$GT_{G_0}$ either contains an infinite number of SLTNF-trees or
has an infinite derivation within some SLTNF-tree. Note that $P$ has only a finite
number of predicate, function and constant symbols. If $GT_{G_0}$ 
contains an infinite number of SLTNF-trees, there must exist negative loops
in $GT_{G_0}$, a contradiction. If $GT_{G_0}$ has an infinite 
derivation within some SLTNF-tree, some positive subgoal $A_0$
selected by $R$ must have infinitely many variant descendants 
$A_1,A_2,..., A_i,...$ on the path 
such that the proof of $A_0$ needs the proof of $A_1$ that needs the proof of 
$A_2$, and so on. That is, $A_i$ is an ancestor loop subgoal of $A_j$ for any
$0\leq i <j$. This contradicts the fact that any descendant loop subgoal in $GT_{G_0}$
has only one ancestor loop subgoal because a descendant loop subgoal cannot
generate descendant loop subgoals since no clauses will be applied to it
for expansion (see case \ref{l4d} of Definition \ref{SLTNF-tree}). $\Box$

\bigskip

Consider Figure \ref{fig2} again. Observe that if we continued 
expanding $N_1$ (like Global SLS-resolution) by applying 
$C_{p_1}$ and $C_{p_2}$, we would generate another correct answer 
substitution $\{Y/c\}$ for $G_0$.
This indicates that applying loop cutting alone
would result in incompleteness.
 
We use {\em answer iteration} \cite{shen2000}
to derive all answers of loop subgoals. Here is the basic idea:
We first build a generalized
SLTNF-tree for $(P\cup \{G_0\}, {\cal T_P}^0)$ with ${\cal T_P}^0 = \emptyset$
while collecting all new tabled answers (for all tabled subgoals) 
into $NEW^0$. Then we build a new generalized
SLTNF-tree for $(P\cup \{G_0\},{\cal T_P}^1)$ with ${\cal T_P}^1 = {\cal T_P}^0 \cup NEW^0$
while collecting all new tabled answers into $NEW^1$. Such an iterative process continues 
until no new tabled answers are available.

The key issue with answer iteration is {\em answer completion}, i.e, how to determine if 
the table of a subgoal is complete at some derivation stages. Careful reader may have
noticed that we have already used a completion criterion for ground subgoals 
in defining the NAF rule (see case \ref{l5d} of Definition \ref{SLTNF-tree}). 
We now generalize this criterion to all subgoals.

\begin{theorem}
\label{comp-thm}
Let $GT_{G_0}$ be the generalized SLTNF-tree for $(P\cup \{G_0\}, \cal T_P)$
and $NEW$ contain all new tabled answers in $GT_{G_0}$.
The following completion criteria hold.
\begin{enumerate}
\item
For a ground tabled positive subgoal $A$, $TB_A\in {\cal T_P} \cup NEW$ 
is complete for $A$ if $TB_A\rightarrow ans = \{A\}$.

\item
For any tabled positive subgoal $A$, $TB_A\in {\cal T_P} \cup NEW$
is complete for $A$ if there is a node $N_i:G_i$ in $GT_{G_0}$, 
where $A$ is the selected subgoal in $G_i$ and let $T_A$ be the sub-SLTNF-tree for $A$ at 
$N_i$, such that (1) $T_A$ has no temporarily undefined leaf, and 
(2) for each loop leaf in $T_A$, the sub-SLTNF-tree for its ancestor loop subgoal $V$
has neither temporarily undefined leaf nor success leaf with a correct answer
substitution $\theta$ such that $V\theta$ is not in $\cal T_P$.
\end{enumerate}
\end{theorem}

\noindent {\bf Proof:}
The first criterion is straightforward since $A$ is ground. We now
prove the second. Note that there are only two cases in which a tabled subgoal $A$
may get new answers via iteration. The first is due to that some temporarily undefined
subgoals in the current round would become successful or failed in the future rounds of 
iteration. This case is excluded by conditions (1) and (2). 
The second case is due to that some loop subgoals in $T_A$ in the current round
would produce new answers in the future rounds of iteration. Such new answers
are generated in an iterative way, i.e., in the current round descendant loop
subgoals in $T_A$ consume only existing tabled answers 
in $\cal T_P$ and help generate new answers
(which are not in $\cal T_P$) for their ancestor loop subgoals. 
These new answers are then tabled (in $NEW$) for the descendant loop subgoals to consume 
in the next round. In this case, $T_A$ must
contain at least one descendant loop subgoal $V'$
such that the sub-SLTNF-tree for its ancestor loop subgoal $V$
has a success leaf with a new correct answer substitution not included 
in $\cal T_P$ (this new answer is not consumed by $V'$ 
in the current round but will be consumed
in the next round). Obviously, this case is excluded by condition (2). 
As a result, conditions (1) and (2) together imply that
further iteration would generate no new answers for $A$. Therefore,
$TB_A$ is complete for $A$ after merging $\cal T_P$ with the new 
tabled answers $NEW$ in $GT_{G_0}$.
$\Box$

\begin{example}
\label{eg3}
{\em
Consider Figure \ref{fig2}. We cannot apply Theorem \ref{comp-thm}
to determine the completeness of $TB_{p(a,Y)}$
since the ancestor loop subgoal $p(a,Y)$ at $N_0$ has a successful sub-derivation
with an answer $p(a,b)$ not in ${\cal T_P}_1$. As we can see, applying this new answer 
to the descendant loop subgoal at $N_1$ would generate another 
new answer $p(a, c)$. The completeness of $TB_s$ is not determinable
either, since both the two sub-SLTNF-trees for $s$ (rooted at $N_6$ and 
$N_8$, respectively) contain a temporarily undefined leaf.
However, by Theorem \ref{comp-thm}, $TB_r$ is complete.
}
\end{example}

\begin{definition}[SLTNF-resolution]
\label{sltnf}
{\em
Let $P$ be a logic program, $G_0 = \leftarrow A$
a top goal with $A$ an atom, and $R$ a computation rule. 
Let ${\cal T_P}^0 = \emptyset$.
{\em SLTNF-resolution} evaluates $G_0$ by calling the function
$SLTNF(P,G_0,R,{\cal T_P}^0)$, defined as follows.
\begin{tabbing}
{\bf function} $SLTNF(P,G_0,R,{\cal T_P}^i)$ {\bf returns} a table $TB_A$\\
$\{$ \\
 $\quad$ \= Build \= a generalized SLTNF-tree $GT_{G_0}^i$ for $(P \cup \{G_0\},{\cal T_P}^i)$ while collecting\\
\> \>                all new tabled answers into $NEW^i$;\\
\> ${\cal T_P}^{i+1} = {\cal T_P}^i\cup NEW^i$;\\ 
\>  Check completeness of all tables in ${\cal T_P}^{i+1}$ and update their status;\\
\> {\bf if} $NEW^i = \emptyset$ or $TB_A \rightarrow comp = 1$ {\bf then return} $TB_A$;\\
\> {\bf return} $SLTNF(P,G_0,R,{\cal T_P}^{i+1})$;\\ 
$\}$
\end{tabbing}
}
\end{definition}  

\begin{figure}[htb]  
\begin{center}
\input{fig3.latex}
\end{center}
\caption{The generalized SLTNF-tree $GT_{\leftarrow p(a,Y)}^1$ for 
$(P_1\cup \{\leftarrow p(a,Y)\},{\cal T_P}_1^1)$.}\label{fig3}
\end{figure}
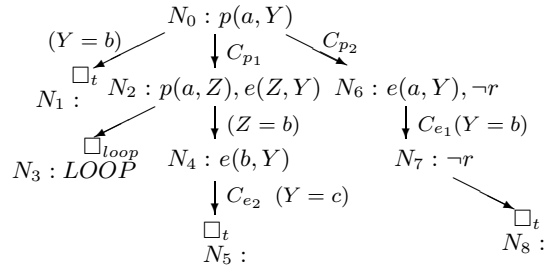

\begin{example}[Cont. of Example \ref{eg3-2}]
{\em
First execute $SLTNF(P_1,G_0,R,{\cal T_P}_1^0)$ where 
${\cal T_P}_1^0=\emptyset$, $G_0 = \leftarrow p(a,Y)$
and $R$ is the left-most computation rule. The procedure 
builds a generalized SLTNF-tree
for $(P_1\cup \{\leftarrow p(a,Y)\},\emptyset)$ as shown 
in Figure \ref{fig2}. It also collects the following
new tabled answer into $NEW^0$: $p(a, b)$ for $TB_{p(a,Y)}$. 
Moreover, it has $TB_r$ completed by setting
$TB_r\rightarrow comp$ to 1 (note that $TB_r\rightarrow ans = \emptyset$).

Next execute $SLTNF(P_1,G_0,R,{\cal T_P}_1^1)$ where 
${\cal T_P}_1^1 = {\cal T_P}_1^0 \cup NEW^0$.
It builds a generalized SLTNF-tree $GT_{\leftarrow p(a,Y)}^1$
for $(P_1\cup \{\leftarrow p(a,Y)\},{\cal T_P}_1^1)$ as shown 
in Figure \ref{fig3}, and collects the following
new tabled answer into $NEW^1$: $p(a, c)$ for $TB_{p(a,Y)}$. 

Finally execute $SLTNF(P_1,G_0,R,{\cal T_P}_1^2)$ 
where ${\cal T_P}_1^2 = {\cal T_P}_1^1 \cup NEW^1$.
The procedure builds a generalized SLTNF-tree $GT_{\leftarrow p(a,Y)}^2$
for $(P_1\cup \{\leftarrow p(a,Y)\},{\cal T_P}_1^2)$ in which no
new tabled answer is produced. Therefore, it returns with two 
tabled answers, $p(a,b)$ and $p(a,c)$, to the top goal $G_0$. 
}
\end{example}

\begin{theorem}
\label{sltnf-term}
Let $P$ be a logic program with the bounded-term-size property, 
$G_0$ a top goal and $R$ a computation rule. $SLTNF(P,G_0,R,\emptyset)$
terminates in finite time.
\end{theorem}

\noindent {\bf Proof:}
Let $n$ be the maximum size of arguments in any top goal. Since $P$ has
the bounded-term-size property, neither subgoals nor tabled answers have
arguments whose size exceeds $f(n)$ for some function $f$. 
Let $s=f(n)$. Since $P$ has a finite number of predicate symbols, 
the number of distinct subgoals (up to
variable renaming) occurring in all $GT_{G_0}^i$s is bounded by a finite number $N(s)$.
Therefore, SLTNF-resolution performs at most $N(s)$ iterations (i.e. generates
at most $N(s)$ generalized SLTNF-trees).
By Theorem \ref{tree-finite}, each iteration 
terminates in finite time, hence SLTNF-resolution 
terminates in finite time. $\Box$

\begin{theorem}
\label{sls-sltnf}
Let $P$ be a logic program with the bounded-term-size property, $A$ an atom,  
and $G_0=\leftarrow A$ a top goal with $A$ a non-floundering query. Let 
$TB_A$ be the tabled answers returned from $SLTNF(P,G_0,R,\emptyset)$, and
let $T_{N_0:G_0}$ be the SLS-tree for $P\cup \{G_0\}$ via $R$.
\begin{enumerate}
\item
$A\theta$ is in $TB_A$ if and only if there is a correct
answer substitution $\theta$ for $G_0$ in $T_{N_0:G_0}$.

\item
$TB_A\rightarrow comp = 1$ and $TB_A\rightarrow ans = \emptyset$  
if and only if $T_{N_0:G_0}$ is failed.
\end{enumerate}
\end{theorem}

\noindent {\bf Proof:}
We first prove that SLS-trees with negative loops can be transformed
into equivalent SLS-trees without negative loops. Let
$T_{N_i:\leftarrow B}$ be an SLS-tree with a descendant 
SLS-tree $T_{N_j:\leftarrow B}$. Obviously, this is a negative loop.
Observe that $B$ at $N_i$ being successful or failed must be independent
of the loop SLS-tree $T_{N_j:\leftarrow B}$, for otherwise the 
truth value of $B$ would depend on $\neg B$ so that $B$ is undefined.
This strongly suggests that using a temporarily undefined value $u^*$
as the truth value of $T_{N_j:\leftarrow B}$ does not change the answer
of $B$ at $N_i$. In other words, any SLS-trees with negative loops can be transformed
into equivalent SLS-trees where all descendant loop SLS-trees are assumed
to return a temporarily undefined value $u^*$.

Let $T_{N_0:G_0}^i$ and $GT_{G_0}^i$ be respectively the SLTNF-tree and  
the generalized SLTNF-tree for $(P\cup \{G_0\}, {\cal T_P}^i)$, where 
${\cal T_P}^0 = \emptyset$ and for each $i\geq 0$, 
${\cal T_P}^{i+1} = {\cal T_P}^i \cup NEW^i$ where $NEW^i$ contains
all new tabled answers collected from $GT_{G_0}^i$.
We prove this theorem by showing that answers over SLS-derivations 
can be extracted in an iterative way and such iterations are the same
as those of SLTNF-resolution. Therefore, both resolutions extract
the same set of answers to $G_0$. We distinguish between three cases:
\begin{enumerate}
\item
For any answer $A\theta$ that is generated without
going through any loops, we must have the same successful derivations for 
$A$ in $T_{N_0:G_0}^0$ as in $T_{N_0:G_0}$. 

\item
\label{pr-2}
Let us consider answers to $G_0$ that are generated without going through any
negative loops. Without loss of generality, assume the SLS-derivations
for the answers involve positive loops as shown in Figure \ref{fig4}, 
where for any $j>k\geq 0$, $B^k$ is
an ancestor loop subgoal of $B^j$ and each $T^k$ together
with the branch leading to
$N_{i^{k+1}}$ is a sub-SLS-tree for $B^k$ at $N_{i^k}$. Obviously, all
$T^k$s are identical up to variable renaming and thus they have the same
set $S_{B^0}$ of correct answer substitutions for 
$B^k$ (up to variable renaming).

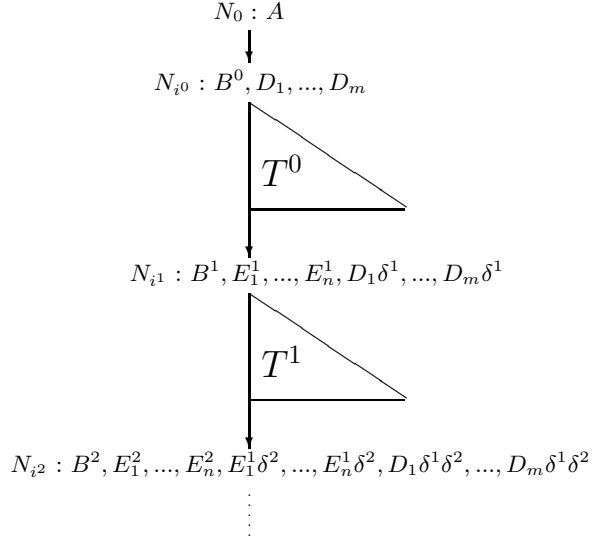
\begin{figure}[htb]  
\begin{center}
\input{fig4.latex}
\end{center}
\caption{SLS-derivations with positive loops.}\label{fig4}
\end{figure} 

Observe that besides $S_{B^0}$, the other possible correct answer substitutions
for $B^k$ must be generated via the infinite loops in an iterative way:
For any $l>0$, the correct answer substitutions for $B^l,E_1^l,...,E_n^l$
at $N_{i^l}$ combined with $\delta^l$, when restricted to the variables in $B^{l-1}$,
are also correct answer substitutions for $B^{l-1}$
at $N_{i^{l-1}}$. These substitutions are obtained by applying each 
correct answer substitution $\theta^l$ for $B^l$ to $E_1^l,...,E_n^l$
and then evaluating $(E_1^l,...,E_n^l)\theta^l$. Since $P$ has the 
bounded-term-size property, no correct answer substitution requires performing
an infinite number of such iterations. That is, there must exist a depth bound $d$ such that
any correct answer substitution $\theta$ for $B^0$ is in the following 
closure (fixpoint):
\begin{itemize}
\item
The initial set of correct answer substitutions
is $S_d = S_{B^0}$.

\item
For each $0<l\leq d$, the set of correct answer substitutions for $B^{l-1}$ at 
$N_{i^{l-1}}$ is
$S_{l-1} = S_l \cup \{\theta |\theta^l\in S_l$ and $\theta = \delta^l\theta^l\alpha$ 
where $\alpha$ is a correct answer substitution for $(E_1^l,...,E_n^l)\theta^l\}$.
\end{itemize}

Apparently, SLTNF-resolution performs the same iterations by making use of the
loop cutting and tabling mechanisms: In the beginning, $TB_{B^0}$
is empty. The loop is cut at $N_{i^1}$, so $TB_{B^0} = S_d = S_{B^0}$
after $T_{N_0:G_0}^0$ is generated (note $B^0$ and 
$B^k$ (resp., $T^0$ and $T^k$) are variants). Then for the $l$-th iteration
$(0<l\leq d)$ $TB_{B^0}$ obtains new answers by applying the already tabled
answers to $B^1$ at $N_{i^1}$ in $T_{N_0:G_0}^l$; i.e., $TB_{B^0} = S_{l-1}$.
As a result, SLS-resolution and SLTNF-resolution derive the same set
of correct answer substitutions for all subgoals involving no negative loops.

\item
Let us now consider answers to $G_0$ that are generated involving
negative loops. As we discussed earlier, loop 
descendant SLS-trees $T_{N_i:\leftarrow B}$
can be removed by assuming they return a temporarily undefined value $u^*$.
Then we get equivalent SLS-trees without any negative loops. By point \ref{pr-2}
above, we can exhaust all answers to $G_0$ from 
these (negative loop free) SLS-trees in an iterative way, as SLTNF-resolution does.
If no single answer to $A$ in $G_0$ is generated after the iteration, we have two
cases. The first is that no SLS-derivation for $A$ at $N_0$ 
ends at a leaf with $u^*$. This means that the truth value of $A$ does not
depend on any negative loop subgoal, so $T_{N_0:G_0}$ is failed 
and thus $TB_A\rightarrow comp = 1$ and $TB_A\rightarrow ans = \emptyset$. 
The second case is that some SLS-derivation for $A$ at $N_0$ 
ends at a leaf with $u^*$. This means that the truth value of $A$
recursively depends on some negative loop subgoal, so $A$ is undefined. In this case,
SLTNF-resolution stops with $TB_A\rightarrow comp = 0$ and 
$TB_A\rightarrow ans = \emptyset$. $\Box$
\end{enumerate}

Since Global SLS-resolution is sound and complete w.r.t. the well-founded 
semantics (see Theorem \ref{sound-comp}), we have the following immediate corollary.

\begin{corollary}
Let $P$ be a logic program, $R$ a computation rule,
and $G_0\leftarrow Q$ be a top goal with $Q$ a non-floundering query under $R$. 
SLTNF-resolution is sound and complete w.r.t. the well-founded semantics.
\end{corollary}

\section{Related Work}
Existing procedural semantics for the well-founded model can be divided 
into two groups in terms of the way they make derivations: (1) {\em bottom-up} approaches, 
such as the alternating fixpoint
approach \cite{VG89-1,lt2001}, the magic sets 
approach \cite{kem95,mor96} and the transformation-based 
bottom-up approach \cite{brass98,brass2001,brass01}, and (2) {\em top-down} approaches. Our method
belongs to the second group. Existing top-down methods can be further
divided into two groups: (1) non-tabling methods, such as Global SLS-resolution,
and (2) tabling methods. Our method is one with tabling. Several tabling methods
for positive logic programs have been proposed, such as OLDT-resolution \cite{TS86},
TP-resolution \cite{shen2000,zhou2000} and the DRA tabling mechanism \cite{guo2001}. However, 
to the best of our knowledge, only SLG-resolution and SLT-resolution use tabling to
compute the well-founded semantics for general logic programs.

SLG-resolution is the state-of-the-art tabling mechanism. It is based on program
transformations, instead of on standard tree-based formulations
like SLDNF- or Global SLS-resolution. Starting from the predicates
of the top goal, it transforms (instantiates) a set of clauses, called a
{\em system}, into another system based on six basic transformation 
rules. Such a system corresponds to a forest of trees with each tree
rooted at a tabled subgoal. A special class of literals, called {\em delaying literals},
is used to represent and handle temporarily undefined negative literals.
Negative loops are identified by maintaining an additional {\em dependency graph}
of subgoals \cite{CSW95, chen96}. In contrast, SLTNF-resolution generates an SLTNF-tree
for the top goal in which the flow of the query evaluation is naturally depicted by the ordered
expansions of tree nodes. Such a tree-style formulation is quite easy for users to understand and 
keep track of the computation. It can also be implemented efficiently using a simple stack-based 
memory structure. The disadvantage of SLTNF-resolution is that
it is a little more costly in time than SLG-resolution due to the use of
answer iteration in exchange for the linearity of derivations.

SLT-resolution is a tabling mechanism with the linearity property. Like SLTNF-resolution,
it expands tree nodes by first applying tabled answers and
then applying clauses. It also uses answer iteration to derive missing
answers caused by loop cuttings. However, it is different from SLTNF-resolution
both in loop handling and in answer completion (note that loop handling 
and answer completion are two key components of a tabling system). 

Recall that SLT-resolution defines positive and negative loops based on the same 
ancestor-descendant relation: Let $A$ be a selected positive subgoal
and $B$ be a subgoal produced by applying a clause to $A$, then $B$
is a descendant subgoal of $A$ and inherits all ancester subgoals of $A$;
let $\neg A$ be a selected ground subgoal with $T_{N_r:\leftarrow A}$ being
its subsidiary SLT-tree, then the subgoal $A$ at $N_r$ inherits 
all ancester subgoals of $\neg A$. A (positive or negative) loop 
occurs when a selected subgoal has an ancestor loop subgoal. Observe that
the ancestor and descendant subgoals may be in different SLT-trees.

When a positive loop occurs, SLTNF-resolution will apply no clauses
to the descendant loop subgoal for node expansion, which guarantees that
any ancestor loop subgoal has just one descendant loop subgoal. However, 
SLT-resolution will continue expanding the
descendant loop subgoal by applying those clauses that have not yet 
been applied by any of its ancestor loop subgoals. As an illustration,
in Figure \ref{fig2}, SLT-resolution will apply $C_{p_2}$ to expand
$N_1$, leading to a child node $N_1'$ with a goal $\leftarrow e(a,Z), \neg r, e(Z, Y)$.
Observe that if the subgoal $e(a,Z)$ at $N_1'$ were $p(a,Z)$, another loop would occur
between $N_0$ and $N_1'$. This suggests that in SLT-resolution, an ancestor loop
subgoal may have several descendant loop subgoals. Due to this, SLT-resolution is
more complicated and costly than SLTNF-resolution in handling positive loops.

SLT-resolution is also more costly than SLTNF-resolution in handling negative loops.
It checks negative loops in the same way as positive loops
by comparing a selected subgoal with all of its ancester subgoals across
all of its ancestor SLT-trees. However, in SLTNF-resolution a negative loop 
is checked simply by comparing a selected ground negative subgoal
with the root goals of its ancestor SLTNF-trees. Recall
that a negative loop occurs if a negative ground subgoal $\neg A$ is selected 
such that the root of the current SLTNF-tree or one of 
its ancestor SLTNF-trees is with a goal $\leftarrow A$. 

SLT-resolution provides no mechanism for answer completion except that 
when a generalized SLT-tree $GT_{G_0}^i$ is generated which contains
no new tabled answers, it evaluates each negative ground
subgoal $\neg A$ in $GT_{G_0}^i$ in a way such that (1) $\neg A$ fails
if $A$ is a tabled answer, and (2) $\neg A$ succeeds if (i)
all branches of its subsidiary SLT-tree $T_{N_r:\leftarrow A}$ end with a failure
leaf and (ii) for each loop subgoal in $T_{N_r:\leftarrow A}$, 
all branches of the sub-SLT-trees for its ancestor loop subgoals 
end with a failure leaf. Not only is this process complicated, it is
also quite inefficient since the evaluation of $\neg A$ may involve
several ancestor SLT-trees. In contrast, SLTNF-resolution provides 
two criteria for completing answers of both negative and positive
subgoals. On the one hand, the criteria
are applied during the construction of generalized SLT-trees so that
redundant derivations can be pruned as early as possible. On
the other hand, checking the completion of a subgoal involves
only one SLTNF-tree.

\section{Conclusions and Further Work}
Global SLS-resolution and SLG-resolution represent two typical styles
in top-down computing the well-founded semantics; the former emphasizes the linearity
of derivations as SLDNF-resolution does while the latter 
focuses on making full use of tabling to resolve
loops and redundant computations. SLTNF-resolution obtains the advantages of
the two methods by enhancing Global SLS-resolution
with loop cutting and tabling mechanisms. It seems that the existing linear
tabling mechanism SLT-resolution has similar advantages, but SLTNF-resolution is 
simpler and more efficient due to its distinct mechanisms for loop handling
and answer completion.

Due to its SLDNF-tree like structure,
SLTNF-resolution can be implemented over a Prolog
abstract machine such as WAM \cite{WAM83} or ATOAM \cite{ZHOU96}. In particular, 
it can be implemented over existing linear tabling systems 
for positive logic programs such as \cite{zhou2000,zhou2003,zhou2004},
simply by adding two more mechanisms, one for identifying negative loops and the other
for checking answer completion of tabled subgoals. We are currently 
working on the implementation. Experimental analysis of SLTNF-resolution will then be reported
in the near future.

\section*{Acknowledgment}
We thank the anonymous referees for their helpful comments.
Yi-Dong Shen is supported in part by Chinese National
Natural Science Foundation and Trans-Century Training Programme
Foundation for the Talents by the Chinese Ministry of Education.

\end{document}

%% file: fig2.latex
\setlength{\unitlength}{3947sp}%
\begingroup\makeatletter\ifx\SetFigFont\undefined%
\gdef\SetFigFont#1#2#3#4#5{%
  \reset@font\fontsize{#1}{#2pt}%
  \fontfamily{#3}\fontseries{#4}\fontshape{#5}%
  \selectfont}%
\fi\endgroup%
\begin{picture}(2862,3846)(2851,-3811)
\thinlines
{\color[rgb]{0,0,0}\put(4351,-136){\vector(-2,-1){380}}
}%
{\color[rgb]{0,0,0}\put(4651,-136){\vector( 2,-1){380}}
}%
{\color[rgb]{0,0,0}\put(5026,-586){\vector( 0,-1){200}}
}%
{\color[rgb]{0,0,0}\put(4421,-1226){\vector(-2,-1){0}}
\multiput(4801,-1036)(-63.33333,-31.66667){6}{\makebox(1.6667,11.6667){\SetFigFont{5}{6}{\rmdefault}{\mddefault}{\updefault}.}}
}%
{\color[rgb]{0,0,0}\put(4426,-1486){\vector( 0,-1){200}}
}%
{\color[rgb]{0,0,0}\put(4426,-1936){\vector( 0,-1){200}}
}%
{\color[rgb]{0,0,0}\put(3671,-2576){\vector(-2,-1){0}}
\multiput(4051,-2386)(-63.33333,-31.66667){6}{\makebox(1.6667,11.6667){\SetFigFont{5}{6}{\rmdefault}{\mddefault}{\updefault}.}}
}%
{\color[rgb]{0,0,0}\put(4426,-2386){\vector( 0,-1){200}}
}%
{\color[rgb]{0,0,0}\put(5326,-1036){\vector( 2,-1){380}}
}%
{\color[rgb]{0,0,0}\put(3751,-611){\vector(-2,-1){380}}
}%
{\color[rgb]{0,0,0}\put(3526,-2836){\vector(-2,-1){380}}
}%
{\color[rgb]{0,0,0}\put(4426,-2836){\vector( 0,-1){200}}
}%
{\color[rgb]{0,0,0}\put(3151,-3286){\vector( 0,-1){200}}
}%
\put(4951,-211){\makebox(0,0)[lb]{\smash{\SetFigFont{8}{9.6}{\rmdefault}{\mddefault}{\updefault}{\color[rgb]{0,0,0}$C_{p_2}$}%
}}}
\put(3826,-211){\makebox(0,0)[lb]{\smash{\SetFigFont{8}{9.6}{\rmdefault}{\mddefault}{\updefault}{\color[rgb]{0,0,0}$C_{p_1}$}%
}}}
\put(4126,-61){\makebox(0,0)[lb]{\smash{\SetFigFont{9}{10.8}{\rmdefault}{\mddefault}{\updefault}{\color[rgb]{0,0,0}$N_0:$ $p(a,Y)$ }%
}}}
\put(4726,-511){\makebox(0,0)[lb]{\smash{\SetFigFont{9}{10.8}{\rmdefault}{\mddefault}{\updefault}{\color[rgb]{0,0,0}$N_3:$  $e(a, Y), \neg r$}%
}}}
\put(5101,-736){\makebox(0,0)[lb]{\smash{\SetFigFont{8}{9.6}{\rmdefault}{\mddefault}{\updefault}{\color[rgb]{0,0,0}$C_{e_1}$}%
}}}
\put(4726,-961){\makebox(0,0)[lb]{\smash{\SetFigFont{9}{10.8}{\rmdefault}{\mddefault}{\updefault}{\color[rgb]{0,0,0}$N_4:$ $\neg r$ }%
}}}
\put(4201,-1411){\makebox(0,0)[lb]{\smash{\SetFigFont{9}{10.8}{\rmdefault}{\mddefault}{\updefault}{\color[rgb]{0,0,0}$N_5:$ $ r$ }%
}}}
\put(4051,-1861){\makebox(0,0)[lb]{\smash{\SetFigFont{9}{10.8}{\rmdefault}{\mddefault}{\updefault}{\color[rgb]{0,0,0}$N_6:$ $s, r$ }%
}}}
\put(3976,-2311){\makebox(0,0)[lb]{\smash{\SetFigFont{9}{10.8}{\rmdefault}{\mddefault}{\updefault}{\color[rgb]{0,0,0}$N_7:$ $\neg s, r$ }%
}}}
\put(5401,-736){\makebox(0,0)[lb]{\smash{\SetFigFont{8}{9.6}{\rmdefault}{\mddefault}{\updefault}{\color[rgb]{0,0,0}$(Y=b)$}%
}}}
\put(5626,-1486){\makebox(0,0)[lb]{\smash{\SetFigFont{9}{10.8}{\rmdefault}{\mddefault}{\updefault}{\color[rgb]{0,0,0}$N_{13}:$}%
}}}
\put(4501,-1636){\makebox(0,0)[lb]{\smash{\SetFigFont{8}{9.6}{\rmdefault}{\mddefault}{\updefault}{\color[rgb]{0,0,0}$C_r$}%
}}}
\put(4501,-2086){\makebox(0,0)[lb]{\smash{\SetFigFont{8}{9.6}{\rmdefault}{\mddefault}{\updefault}{\color[rgb]{0,0,0}$C_s$}%
}}}
\put(5701,-1336){\makebox(0,0)[lb]{\smash{\SetFigFont{9}{10.8}{\rmdefault}{\mddefault}{\updefault}{\color[rgb]{0,0,0}$\Box_t$}%
}}}
\put(3001,-511){\makebox(0,0)[lb]{\smash{\SetFigFont{9}{10.8}{\rmdefault}{\mddefault}{\updefault}{\color[rgb]{0,0,0}$N_1:$  $p(a, Z), e(Z,Y)$}%
}}}
\put(3376,-2761){\makebox(0,0)[lb]{\smash{\SetFigFont{9}{10.8}{\rmdefault}{\mddefault}{\updefault}{\color[rgb]{0,0,0}$N_8:$ $s$ }%
}}}
\put(4276,-2761){\makebox(0,0)[lb]{\smash{\SetFigFont{9}{10.8}{\rmdefault}{\mddefault}{\updefault}{\color[rgb]{0,0,0}$N_{11}:$ $r, u^*$ }%
}}}
\put(3301,-961){\makebox(0,0)[lb]{\smash{\SetFigFont{9}{10.8}{\rmdefault}{\mddefault}{\updefault}{\color[rgb]{0,0,0}$\Box_{loop}$}%
}}}
\put(4351,-3211){\makebox(0,0)[lb]{\smash{\SetFigFont{9}{10.8}{\rmdefault}{\mddefault}{\updefault}{\color[rgb]{0,0,0}$\Box_{loop}$}%
}}}
\put(3076,-1111){\makebox(0,0)[lb]{\smash{\SetFigFont{9}{10.8}{\rmdefault}{\mddefault}{\updefault}{\color[rgb]{0,0,0}$N_2: LOOP$}%
}}}
\put(4201,-3361){\makebox(0,0)[lb]{\smash{\SetFigFont{9}{10.8}{\rmdefault}{\mddefault}{\updefault}{\color[rgb]{0,0,0}$N_{12}: LOOP$ }%
}}}
\put(2851,-3211){\makebox(0,0)[lb]{\smash{\SetFigFont{9}{10.8}{\rmdefault}{\mddefault}{\updefault}{\color[rgb]{0,0,0}$N_9:$ $\neg s$ }%
}}}
\put(3076,-2911){\makebox(0,0)[lb]{\smash{\SetFigFont{8}{9.6}{\rmdefault}{\mddefault}{\updefault}{\color[rgb]{0,0,0}$C_s$}%
}}}
\put(3076,-3661){\makebox(0,0)[lb]{\smash{\SetFigFont{9}{10.8}{\rmdefault}{\mddefault}{\updefault}{\color[rgb]{0,0,0}$\Box_{u^*}$}%
}}}
\put(3001,-3811){\makebox(0,0)[lb]{\smash{\SetFigFont{9}{10.8}{\rmdefault}{\mddefault}{\updefault}{\color[rgb]{0,0,0}$N_{10}: u^*$}%
}}}
\end{picture}

%% file: fig3.latex
\setlength{\unitlength}{3947sp}%
\begingroup\makeatletter\ifx\SetFigFont\undefined%
\gdef\SetFigFont#1#2#3#4#5{%
  \reset@font\fontsize{#1}{#2pt}%
  \fontfamily{#3}\fontseries{#4}\fontshape{#5}%
  \selectfont}%
\fi\endgroup%
\begin{picture}(3162,1596)(3301,-1561)
\thinlines
{\color[rgb]{0,0,0}\put(5026,-136){\vector( 2,-1){380}}
}%
{\color[rgb]{0,0,0}\put(4576,-586){\vector( 0,-1){200}}
}%
{\color[rgb]{0,0,0}\put(4576,-1061){\vector( 0,-1){200}}
}%
{\color[rgb]{0,0,0}\put(4201,-586){\vector(-2,-1){380}}
}%
{\color[rgb]{0,0,0}\put(4576,-136){\vector( 0,-1){200}}
}%
{\color[rgb]{0,0,0}\put(4261,-121){\vector(-2,-1){444}}
}%
{\color[rgb]{0,0,0}\put(5776,-586){\vector( 0,-1){200}}
}%
{\color[rgb]{0,0,0}\put(6076,-1036){\vector( 2,-1){380}}
}%
\put(5326,-511){\makebox(0,0)[lb]{\smash{\SetFigFont{9}{10.8}{\rmdefault}{\mddefault}{\updefault}{\color[rgb]{0,0,0}$N_6:$  $e(a, Y), \neg r$}%
}}}
\put(4501,-1411){\makebox(0,0)[lb]{\smash{\SetFigFont{9}{10.8}{\rmdefault}{\mddefault}{\updefault}{\color[rgb]{0,0,0}$\Box_t$}%
}}}
\put(4501,-1561){\makebox(0,0)[lb]{\smash{\SetFigFont{9}{10.8}{\rmdefault}{\mddefault}{\updefault}{\color[rgb]{0,0,0}$N_5:$}%
}}}
\put(3751,-886){\makebox(0,0)[lb]{\smash{\SetFigFont{9}{10.8}{\rmdefault}{\mddefault}{\updefault}{\color[rgb]{0,0,0}$\Box_{loop}$}%
}}}
\put(3676,-436){\makebox(0,0)[lb]{\smash{\SetFigFont{9}{10.8}{\rmdefault}{\mddefault}{\updefault}{\color[rgb]{0,0,0}$\Box_t$}%
}}}
\put(5251,-211){\makebox(0,0)[lb]{\smash{\SetFigFont{8}{9.6}{\rmdefault}{\mddefault}{\updefault}{\color[rgb]{0,0,0}$C_{p_2}$}%
}}}
\put(4651,-286){\makebox(0,0)[lb]{\smash{\SetFigFont{8}{9.6}{\rmdefault}{\mddefault}{\updefault}{\color[rgb]{0,0,0}$C_{p_1}$}%
}}}
\put(3301,-1036){\makebox(0,0)[lb]{\smash{\SetFigFont{9}{10.8}{\rmdefault}{\mddefault}{\updefault}{\color[rgb]{0,0,0}$N_3: LOOP$}%
}}}
\put(4276,-961){\makebox(0,0)[lb]{\smash{\SetFigFont{9}{10.8}{\rmdefault}{\mddefault}{\updefault}{\color[rgb]{0,0,0}$N_4:$  $e(b, Y)$}%
}}}
\put(4651,-1186){\makebox(0,0)[lb]{\smash{\SetFigFont{8}{9.6}{\rmdefault}{\mddefault}{\updefault}{\color[rgb]{0,0,0}$C_{e_2}$}%
}}}
\put(4276,-61){\makebox(0,0)[lb]{\smash{\SetFigFont{9}{10.8}{\rmdefault}{\mddefault}{\updefault}{\color[rgb]{0,0,0}$N_0:$ $p(a,Y)$ }%
}}}
\put(3901,-511){\makebox(0,0)[lb]{\smash{\SetFigFont{9}{10.8}{\rmdefault}{\mddefault}{\updefault}{\color[rgb]{0,0,0}$N_2:$  $p(a, Z), e(Z,Y)$}%
}}}
\put(3451,-586){\makebox(0,0)[lb]{\smash{\SetFigFont{9}{10.8}{\rmdefault}{\mddefault}{\updefault}{\color[rgb]{0,0,0}$N_1:$}%
}}}
\put(5851,-736){\makebox(0,0)[lb]{\smash{\SetFigFont{8}{9.6}{\rmdefault}{\mddefault}{\updefault}{\color[rgb]{0,0,0}$C_{e_1}$}%
}}}
\put(6376,-1486){\makebox(0,0)[lb]{\smash{\SetFigFont{9}{10.8}{\rmdefault}{\mddefault}{\updefault}{\color[rgb]{0,0,0}$N_8:$}%
}}}
\put(6451,-1336){\makebox(0,0)[lb]{\smash{\SetFigFont{9}{10.8}{\rmdefault}{\mddefault}{\updefault}{\color[rgb]{0,0,0}$\Box_t$}%
}}}
\put(5701,-961){\makebox(0,0)[lb]{\smash{\SetFigFont{9}{10.8}{\rmdefault}{\mddefault}{\updefault}{\color[rgb]{0,0,0}$N_7:$ $\neg r$ }%
}}}
\put(4951,-1186){\makebox(0,0)[lb]{\smash{\SetFigFont{8}{9.6}{\rmdefault}{\mddefault}{\updefault}{\color[rgb]{0,0,0}$(Y=c)$}%
}}}
\put(4651,-736){\makebox(0,0)[lb]{\smash{\SetFigFont{8}{9.6}{\rmdefault}{\mddefault}{\updefault}{\color[rgb]{0,0,0}$(Z=b)$}%
}}}
\put(3526,-211){\makebox(0,0)[lb]{\smash{\SetFigFont{8}{9.6}{\rmdefault}{\mddefault}{\updefault}{\color[rgb]{0,0,0}$(Y=b)$}%
}}}
\put(6076,-736){\makebox(0,0)[lb]{\smash{\SetFigFont{8}{9.6}{\rmdefault}{\mddefault}{\updefault}{\color[rgb]{0,0,0}$(Y=b)$}%
}}}
\end{picture}

%% file: fig4.latex
\setlength{\unitlength}{3947sp}%
\begingroup\makeatletter\ifx\SetFigFont\undefined%
\gdef\SetFigFont#1#2#3#4#5{%
  \reset@font\fontsize{#1}{#2pt}%
  \fontfamily{#3}\fontseries{#4}\fontshape{#5}%
  \selectfont}%
\fi\endgroup%
\begin{picture}(2487,3367)(2851,-3323)
\thinlines
{\color[rgb]{0,0,0}\multiput(4351,-3061)(0.00000,-62.50000){5}{\makebox(1.6667,11.6667){\SetFigFont{5}{6}{\rmdefault}{\mddefault}{\updefault}.}}
}%
{\color[rgb]{0,0,0}\put(4351,-586){\vector( 0,-1){975}}
}%
{\color[rgb]{0,0,0}\put(4351,-1786){\vector( 0,-1){975}}
}%
{\color[rgb]{0,0,0}\put(4351,-586){\line( 3,-2){986.538}}
\put(5326,-1261){\line(-1, 0){975}}
}%
{\color[rgb]{0,0,0}\put(4351,-1786){\line( 3,-2){986.538}}
\put(5326,-2461){\line(-1, 0){975}}
}%
{\color[rgb]{0,0,0}\put(4351,-136){\vector( 0,-1){200}}
}%
\put(4126,-61){\makebox(0,0)[lb]{\smash{\SetFigFont{9}{10.8}{\rmdefault}{\mddefault}{\updefault}{\color[rgb]{0,0,0}$N_0:$ $A$ }%
}}}
\put(3751,-511){\makebox(0,0)[lb]{\smash{\SetFigFont{9}{10.8}{\rmdefault}{\mddefault}{\updefault}{\color[rgb]{0,0,0}$N_{i^0}:$ $ B^0,D_1, ..., D_m$ }%
}}}
\put(3601,-1711){\makebox(0,0)[lb]{\smash{\SetFigFont{9}{10.8}{\rmdefault}{\mddefault}{\updefault}{\color[rgb]{0,0,0}$N_{i^1}:$ $ B^1,E_1^1,...,E_n^1,D_1\delta^1, ..., D_m\delta^1$ }%
}}}
\put(2851,-2911){\makebox(0,0)[lb]{\smash{\SetFigFont{9}{10.8}{\rmdefault}{\mddefault}{\updefault}{\color[rgb]{0,0,0}$N_{i^2}:$ $ B^2,E_1^2, ..., E_n^2, E_1^1\delta^2,...,E_n^1\delta^2,D_1\delta^1\delta^2, ..., D_m\delta^1\delta^2$ }%
}}}
\put(4426,-1111){\makebox(0,0)[lb]{\smash{\SetFigFont{14}{16.8}{\rmdefault}{\mddefault}{\updefault}{\color[rgb]{0,0,0}$T^0$}%
}}}
\put(4426,-2311){\makebox(0,0)[lb]{\smash{\SetFigFont{14}{16.8}{\rmdefault}{\mddefault}{\updefault}{\color[rgb]{0,0,0}$T^1$}%
}}}
\end{picture}